\documentclass{iopart}
\expandafter\let\csname equation*\endcsname\relax
\expandafter\let\csname endequation*\endcsname\relax
\usepackage{amsmath, amssymb, amsfonts}
\usepackage{color}
\usepackage{subfig}
\usepackage{graphicx}
\graphicspath{{./}{figs/}}
\newcommand{\figref}[1]{Fig:\,\ref{#1}}
\newcommand{\tabref}[1]{Table:\,\ref{#1}}
\newcommand{\fm}{\thinspace{\textrm{fm}}}     
\newcommand{\smax}{_{\textrm{\tiny{max}}}}    
\newcommand{\mn}{_{m,n}} 

\newcommand{\AGeV}{\,\textrm{AGeV}}
\newcommand{\cR}{\mathcal{R}}
\newcommand{\Na}{{N_a}} 
\newcommand{\Nf}{{N_f}} 

\begin{document}


\title{Classification of initial state granularity via 2d Fourier expansion}

\author{C.E.~Coleman-Smith$^1$,H.~Petersen$^2$, R.L.~Wolpert$^3$}
\ead{cec24@phy.duke.edu}
\address{$^1$ Department of Physics, Duke University, Durham, NC 27708-0305, USA} 
\address{$^2$ Department of Physics, Duke University, Durham, NC 27708-0305, USA} 
\address{$^2$ Frankfurt Institute for Advanced Studies, 
  60438 Frankfurt am Main, Germany} 
\address{$^3$Department of Statistical Science, Duke University, Durham, NC
  27708-0251, USA}. 
\date{\today}

\begin{abstract}

  A new method to quantify fluctuations in the initial state of heavy ion
  collisions is presented.  The initial state energy distribution is
  decomposed with a set of orthogonal basis functions which include both
  angular and radial variation.  The resulting two dimensional Fourier
  coefficients provide additional information about the nature of the
  initial state fluctuations compared to a purely angular decomposition.
  We apply this method to ensembles of initial states generated by both
  Glauber and Color Glass Condensate Monte-Carlo codes.  In addition
  initial state configurations with varying amounts of fluctuations
  generated by a dynamic transport approach are analyzed to test the
  sensitivity of the procedure.  The results allow for a full
  characterization of the initial state structures that is useful to
  discriminate the different initial state models currently in use.

\end{abstract}


\submitto{\JPG}


\maketitle


Ultra-Relativistic nearly-ideal fluid dynamics has proven to be a very
successful tool for modeling the bulk dynamics of the hot dense matter
formed during a heavy ion collision ~\cite {Song:2010mg, Nonaka:2006yn,
  Huovinen:2001cy, Schenke:2010nt, Huovinen:2005gy, Hirano:2002ds}.  The
major uncertainty in determining transport properties of the QGP, such as
the ratio of shear viscosity to entropy, lies in the specification of the
initial conditions of the collision.  The initial conditions have been
mainly assumed to be smooth distributions that are parametrized
implementations of certain physical assumptions (e.g., Glauber/CGC).
Within the last 2 years the importance of including fluctuations in these
distributions has been recognized, leading to a whole new set of
experimental observations of higher flow coefficients and their
correlations ~\cite{Qiu:2011hf, Schenke:2010rr, Alver:2010dn,
  Teaney:2010vd}.  On the theoretical side there has been a lot of effort
to refine the previously schematic models with fluctuation inducing
corrections and to employ adynamical descriptions of the early
non-equilibrium evolution ~\cite{Schenke:2012wb, Qin:2010pf,Dumitru:2012yr,
  Alvioli:2011sk}.

Hydrodynamical simulations can take these fluctuations into account by
generating an ensemble of runs each with a unique initial condition,
so-called \emph {event by event} simulations.  This is in contrast to \emph
{event averaged} simulations where an ensemble of fluctuating initial
conditions is generated, and then a single initial condition corresponding
to this set's ensemble average is subject to evolution.  Event by event
modeling has proven to be essential for correctly describing all the
details of the bulk behavior of heavy ion collisions ~\cite{Qiu:2011iv,
  Staig:2011wj, Bleicher:1998wd, Grassi:2005pm, Tavares:2007mu,
  Andrade:2006yh, Andrade:2008xh, Andrade:2010xy, Holopainen:2010gz,
  Werner:2010aa, Petersen:2010cw}.

The two main models for the generation of hydrodynamic initial conditions
are the Glauber ~\cite{Glauber:1970, Miller:2007ri, Esumi:2011nd,
  Qin:2010pf} and color glass condensate (CGC) models ~\cite{KLN:1, KLN:2,
  Gelis:2010nm, Kovner:1995ts, Krasnitz:1999wc, McLerran:1994vd}.  The
Glauber model samples a Woods-Saxon nuclear density distribution for each
nucleus.

Color glass condensate models are \emph{ab initio} calculations motivated
by the idea of gluon saturation of parton distribution functions at small
momentum scales $x$.  In CGC models the gluon distribution for each nucleon
is computed and the nuclear collision is modeled as interactions between
these coherent color fields.  Each of these models generates spatial
fluctuations whose details depend on the assumptions made in the
specific implementation.  Glauber fluctuations come from Monte-Carlo (MC)
sampling the nuclear density distribution.  CGC fluctuations arise
similarly with additional contributions from the self interaction of the
color fields.

We present a method for generating a $2d$ decomposition of
fluctuations in the initial state energy density of a heavy-ion
collision.  We apply this framework to the ensembles of initial states
generated by UrQMD ~\cite{Bass:1998ca, Bleicher:1999xi,
  Petersen:2008dd}, by an MC Glauber code of Qin ~\cite{Qin:2010pf},
and by the MC-KLN code of Drescher and Nara ~\cite{KLN:1, KLN:2,
  Drescher:2006ca} which is a based on CGC ideas.  The events compared
are generated for Au+Au collisions at $\sqrt{s} = 200\AGeV$ at two
impact parameters $b=2,7\fm$.  We introduce summary statistics for the
Fourier expansion and show how the radial information exposes clear
differences between the fluctuations generated by Glauber and CGC
based codes. We note that the cumulant expansion of Teaney and Yan
\cite{Teaney:2010vd} provides an alternative $2d$ decomposition of the
initial state density. The cumulant basis used therein has many
appealing properties and the familiar $1d$ moments can be readily
obtained from it. The norms wwhich arise from the expansion we propose
below have similar invariance properties and as we shall show very
naturally reveal the underlying roughness of the event.

Since colliding nuclei are strongly Lorentz contracted along the beam axis,
we neglect the longitudinal dimension.  A priori we expect fluctuations in
both radial and azimuthal directions.  Experiments make measurements in a
$2d$ transverse momentum space, and further analysis of the final state may
reveal correlations with quantities derived from a fully $2d$ decomposition
of the initial state.  Recent work \cite{Gardim:2011re} has shown that
events can be constructed which have identical angular Fourier moments
$\epsilon_2, \epsilon_3$ but dramatically different energy distributions
leading to different final state flow coefficients.  The structure of such
events is not sensitive to a purely azimuthal decomposition.

  We seek a two dimensional Fourier expansion on a disk of radius $r_0>0$.
  The orthogonality of Bessel functions of the first kind is the key to
  this decomposition
\begin{equation}
  \label{eqn-ortho-besselJ}
  \int_0^{r_0} J_\alpha\big(\frac{r}{r_0} \lambda_{\alpha, n}\big)
              J_\alpha\big(\frac{r}{r_0} \lambda_{\alpha, n'}\big) r dr
   = \frac{r_0^2}{2} \delta_{nn'}[J_{\alpha+1}(\lambda_{\alpha, n})]^2,
      \quad \forall n,n' \in \mathbb{Z}, \forall \alpha \in \mathbb{R},
\end{equation}

where $\lambda_{\alpha,n}$ is the $n^{th}$ positive zero of
$J_{\alpha}(x)$.  For integer $\alpha$, $J_{\alpha}(-x) =
(-1)^{\alpha}J_{\alpha}(x)$ and so $[J_{\alpha+1}(\lambda_{\alpha,n})]^2 =
[J_{|\alpha|+1}(\lambda_{\alpha,n})]^2$ and $\lambda_{\alpha,n} =
\lambda_{-\alpha,n}$.  It follows that the functions
\begin{equation}
  \label{eqn-basis-fn}
  \phi\mn(r,\theta) := \frac{1}{J_{|m|+1}\big(\lambda\mn\big)}
                        J_m\big(\frac{r}{r_0} \lambda\mn\big) e^{im\theta}
\end{equation}
form a complete orthonormal set on this disk (under the uniform measure
$r\,dr\,d\theta/\pi r_0^2$).  As such any well behaved (square-integrable and
vanishing at the boundary of the disk) function $f$ on this disk admits the
convergent expansion
\begin{equation} \label{eqn-fourier-series}
  f(r,\theta) = \sum\mn A\mn \phi\mn(r, \theta),
\end{equation}
in terms of these basis functions, with generalized Fourier coefficients
$A\mn \in \mathbb{C}$ given by
\begin{equation} \label{eqn-fourier-coeff}
  A\mn = \frac{1}{\pi r_0^2}
         \int f(r,\theta) \phi^{\star}\mn(r, \theta) r dr d\theta.
\end{equation}

A simple scaling law describes the dependence of coefficients $A\mn$
on $r_0$, which thus is arbitrary so long as the support of $f$ is
contained in the ball of radius $r_0$. The parameter $r_0$ sets the
maximum radial position of features in the event that will be resolved
in the decomposition. Throughout the paper we use $r_0 = 10\fm$,
sufficient for all the distributions we consider.  We center the
coordinate system at the center of mass of the distributions $f$.

\begin{figure}
  \centering
  \includegraphics[width=0.35\textwidth]{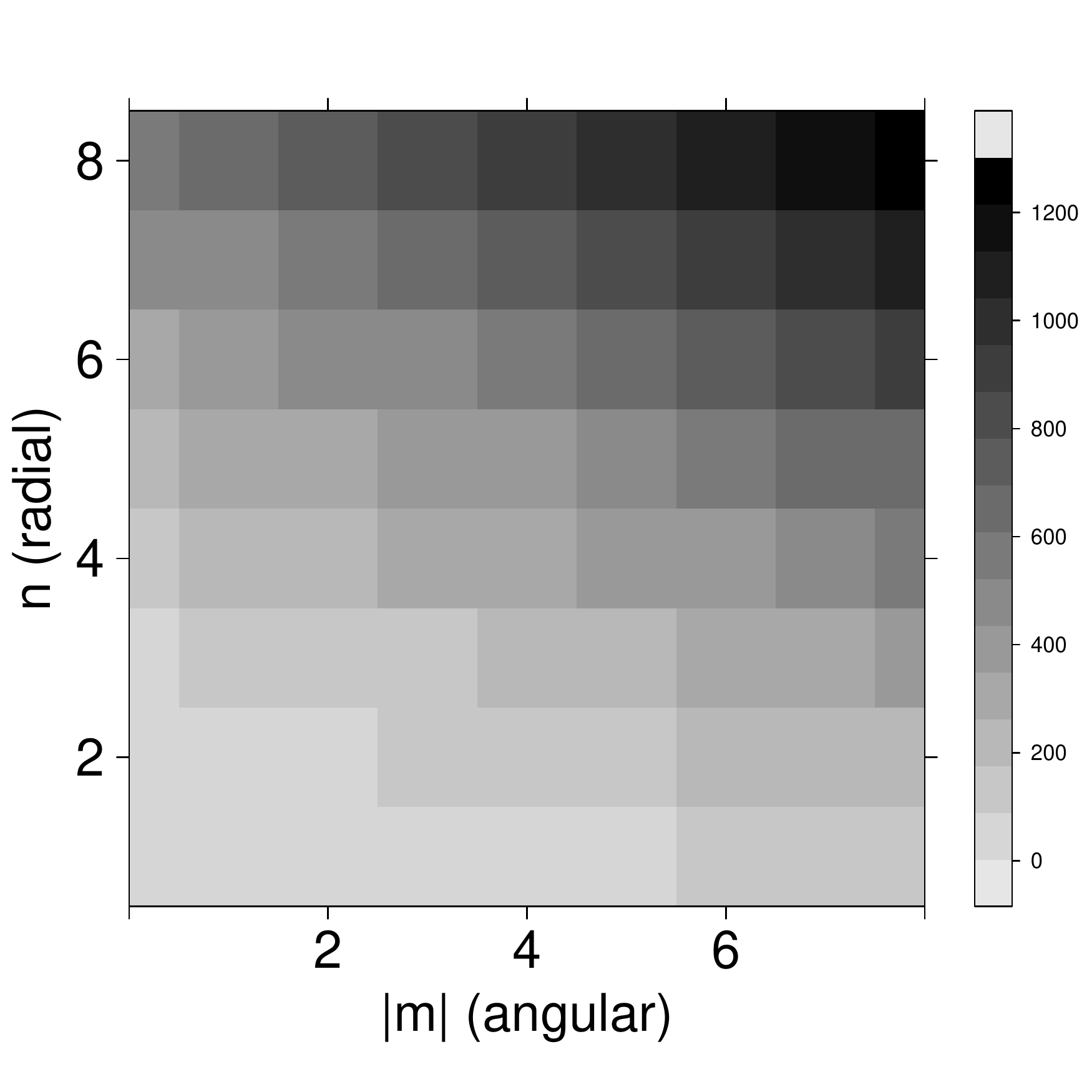}
  \caption{The eigenvalues $\lambda\mn^2$ of the negative Dirichlet
    Laplacian on the unit disk are plotted as a function of $|m|$ (angular)
    and $n$ (radial).  These can be interpreted as an inverse length scale
    for the coefficients $A\mn$.  Asymptotically $|\lambda
    \mn|\asymp\pi(n+|m|-1/4)$ as $n,m\to\infty$.}
  \label{fig-eigenvalues}
\end{figure}

The basis functions $\phi\mn$ are solutions to Bessel's equation on the
unit disk \cite{Abramowitz:1972}, eigenfunctions of the Laplacian with
Dirichlet bc and eigenvalues $-(\lambda\mn/r_0)^2$.  Thus for sufficiently
smooth $f$,
\[
   -\nabla^2 f(r, \theta) = \sum\mn \frac{\lambda\mn^2}{r_0^2} A\mn \phi\mn.
\]
The values $r_0/\lambda\mn$ constitute characteristic length scales for the
associated Fourier components $A\mn$.  Higher orders of $m$ and $n$ are
associated with larger $\lambda\mn$ (see \figref{fig-eigenvalues}),
corresponding to shorter length scales, and are associated with
smaller-scale features (see \cite {Mocsy:2010um, Staig:2010pn} for a
similar perspective in 1d).  Other boundary conditions (Neumann, for
example) lead to similar countable sets of basis functions.

\begin{figure}[htb]
  \centering
    \includegraphics[width=0.45\textwidth, clip, trim=0 0 0 0] 
                    {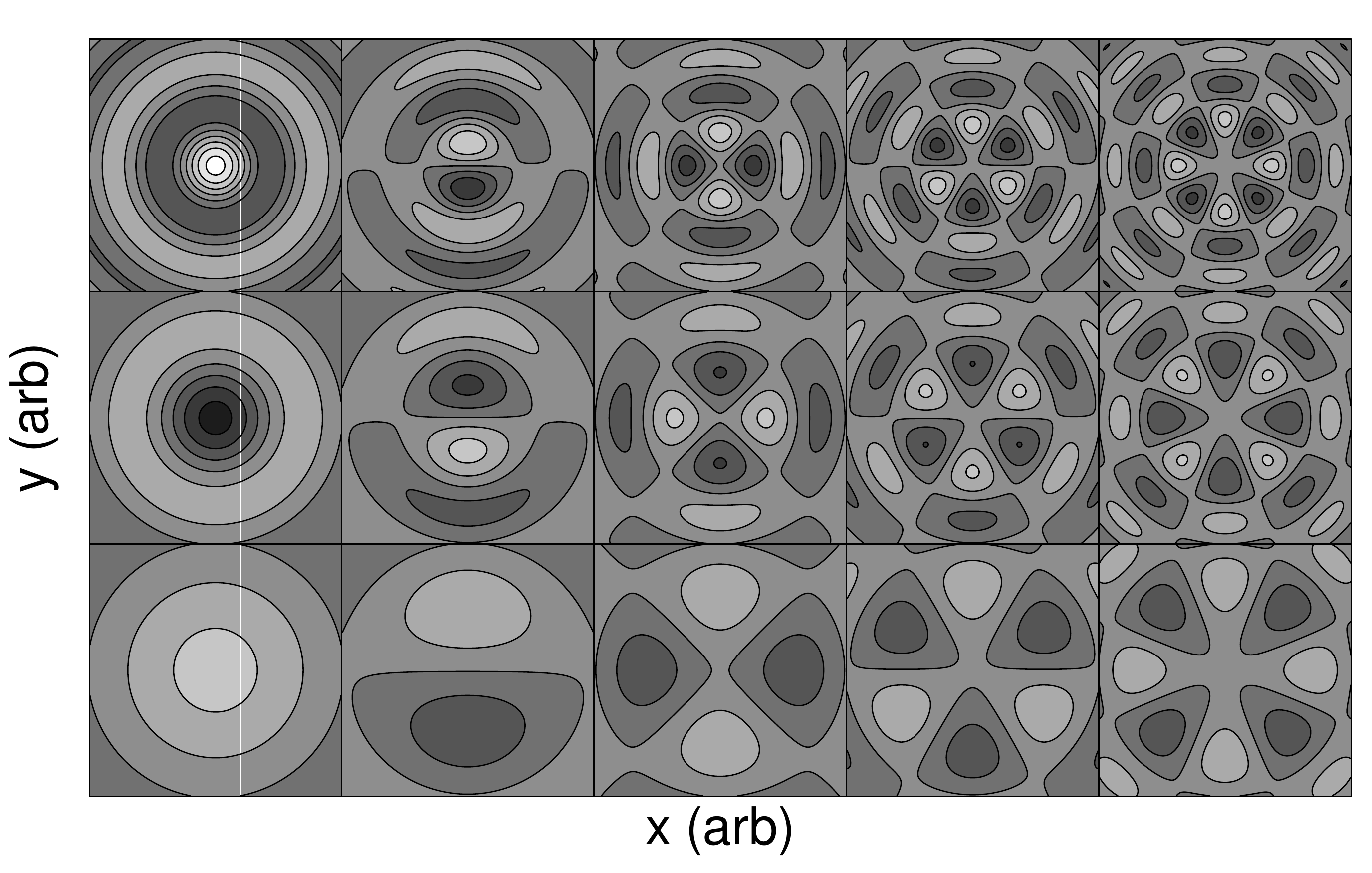}
    \includegraphics[width=0.45\textwidth, clip, trim=0 0 0 0]
                    {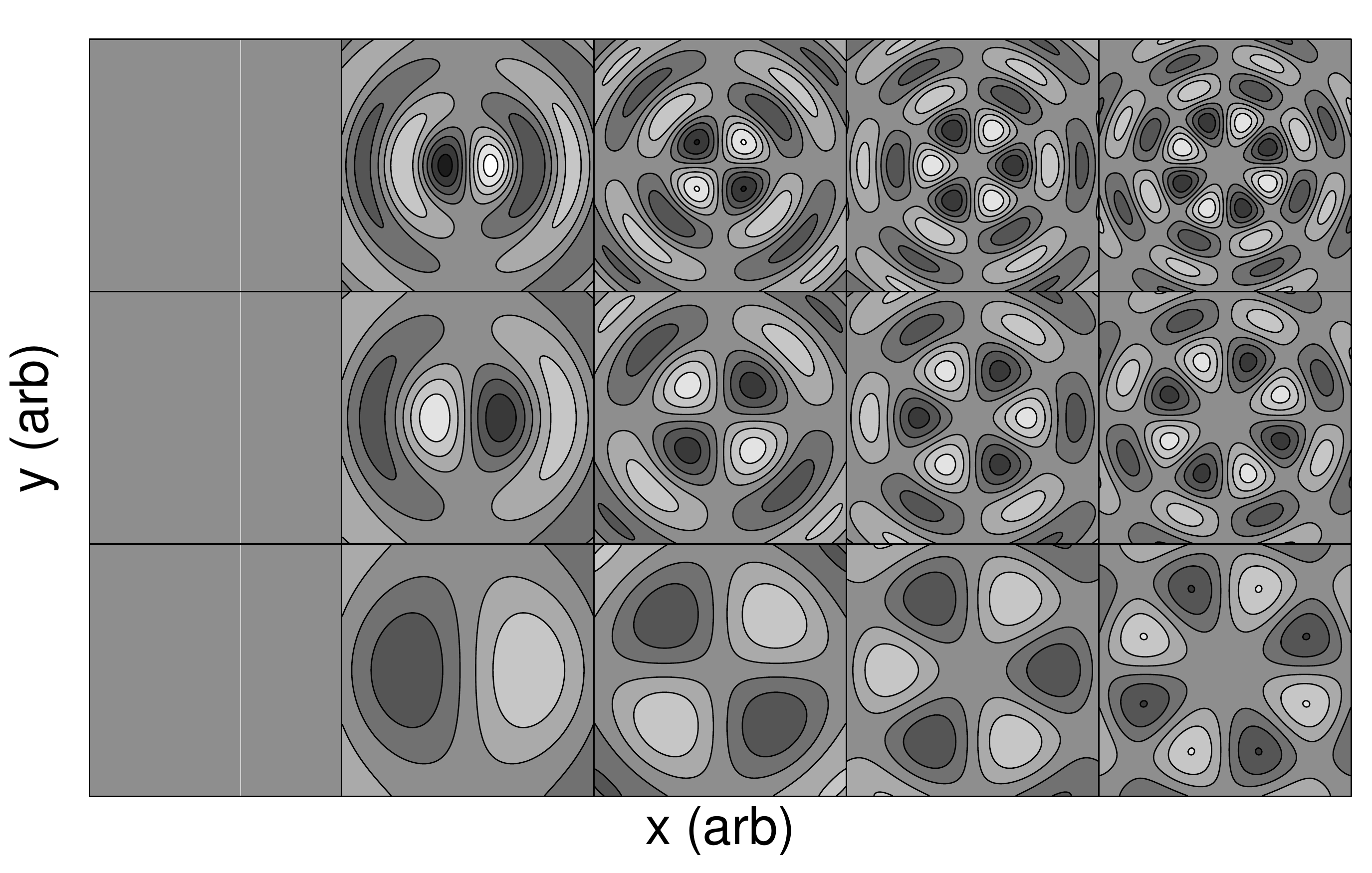}
    \caption{Plots of the first few real (left) and imaginary (right)
      components of $\phi\mn(r,\theta)$.  The angular coefficient $m \in
      [0, 5]$ increases from left to right, the radial coefficient $n \in
      [1,3]$ increases from bottom to top.} 
      \label{fig-basis-fns} 
\end{figure}

Higher values of the angular indices $\pm m$ correspond to higher numbers of
zero crossings in the angular components of basis functions, or ``lumpiness''
in rotation, while higher values of the radial indices $n$ are associated
with more roughness as one moves closer or farther from the center of mass.
The first few basis functions are plotted in \figref{fig-basis-fns}.  The
lumpy shape of the basis functions suggests that this representation will be
very useful for characterizing the hot and cold spot structures in the
initial state of a heavy ion reaction.

\begin{figure*}[htb]
  \centering
    \includegraphics[width=0.30\textwidth, clip, trim=0.5mm 5mm 10mm 10mm]
                    {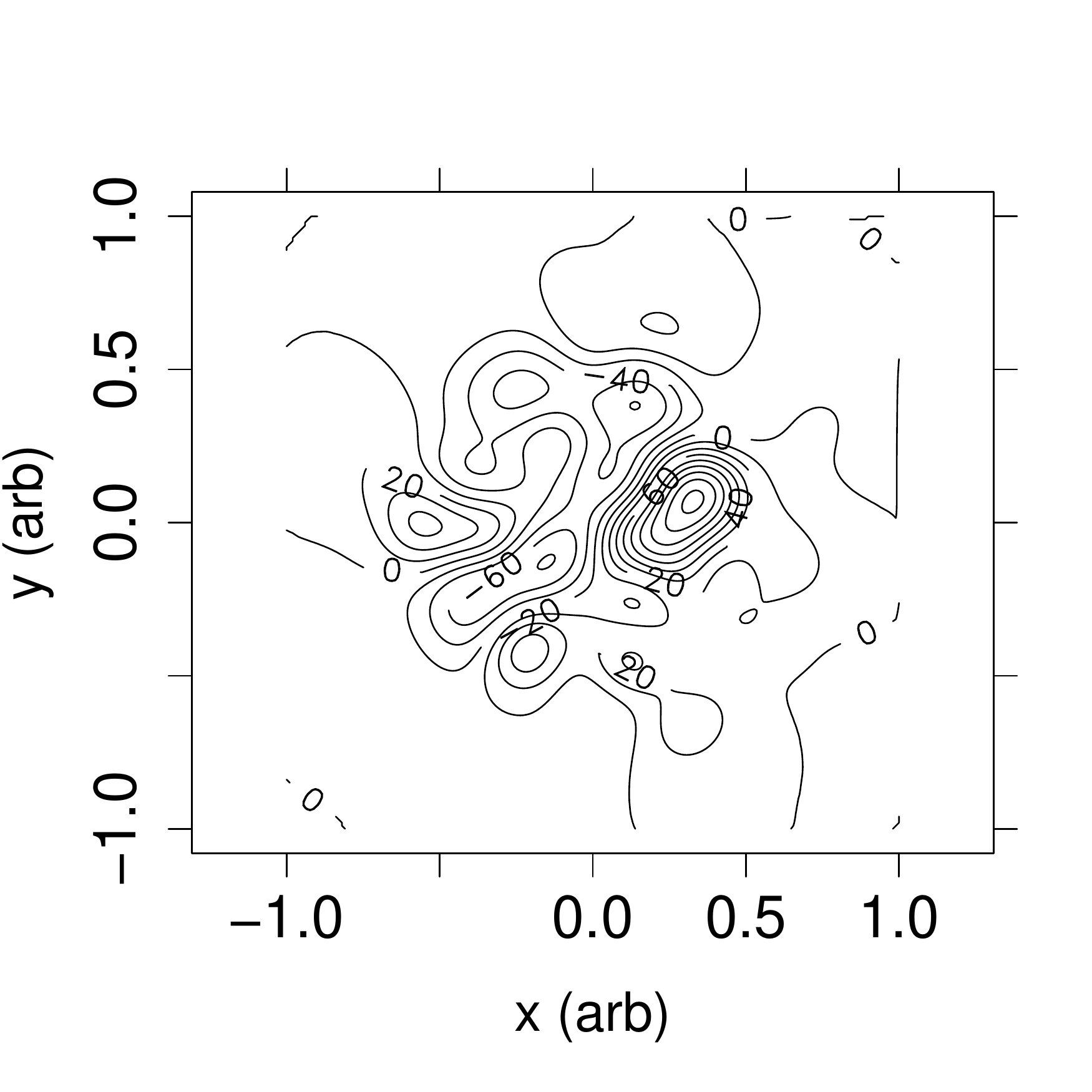}
    \includegraphics[width=0.30\textwidth, clip, trim=0.5mm 5mm 10mm 15mm]
                    {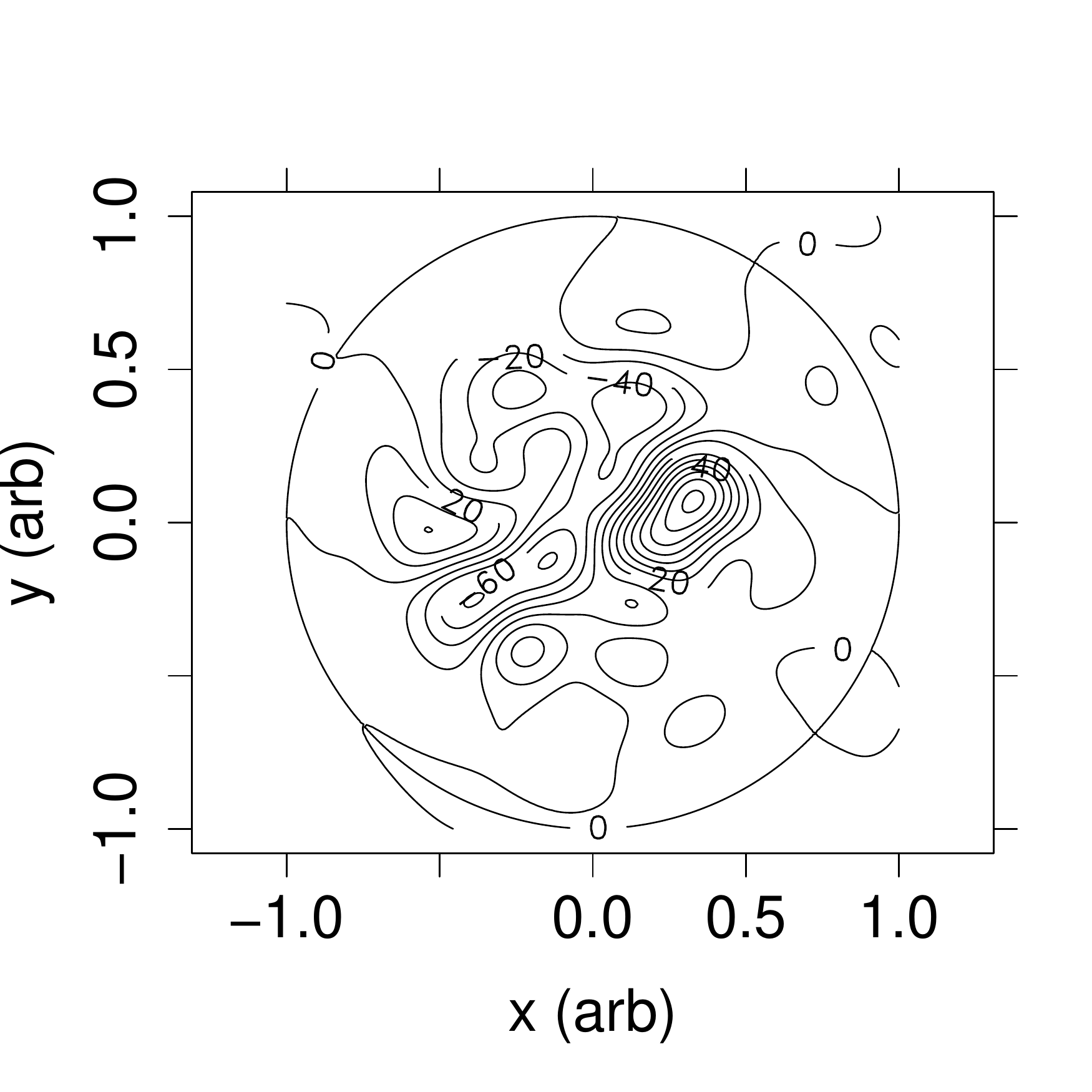}
    \includegraphics[width=0.28\textwidth, clip, trim=0 7.5mm 5mm 5mm]
                    {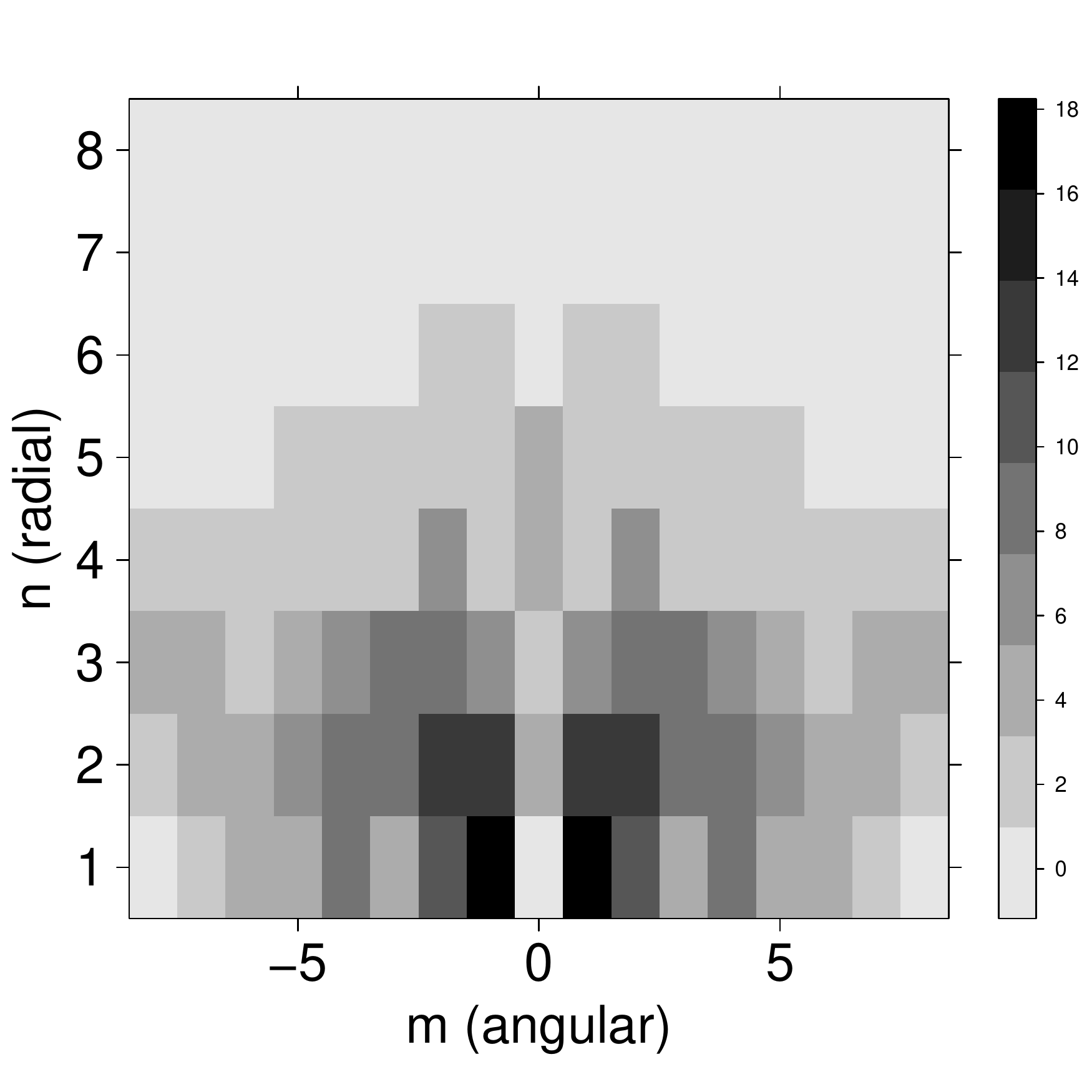}
  \caption{First: A typical UrQMD ensemble average subtracted event.  Second: the
    reconstructed event generated by \eqref{eqn-fourier-series} after
    applying the decomposition.  Third: The absolute values of the Fourier
    coefficients $|A\mn|$ for this event.}
\label{fig-decomp-example}
\end{figure*}

A typical UrQMD event (after subtracting the ensemble average) is shown in
\figref{fig-decomp-example} alongside its decomposition and the resulting
Fourier coefficients $A\mn$.  Here we characterize the event in terms of
coefficients $m \in [ -8, 8]$ and $n \in [1, 8]$, sufficient to capture a
detailed image of the original initial state distribution (further terms
contribute very little additional information because differences between
the original and reconstructed distributions are dominated by numerical
noise for coefficients with order higher than $|m|\smax=8, n\smax=8$).  Let
us now explore different ways to extract useful information from this
decomposition.

Using the orthogonality of the basis functions along with the Laplacian we
can derive simple expressions for norms of the function $f$ to be expanded in
the frequency domain.  The $L_2(f)$ norm, a measure of the total mass of $f$,
is given by Plancherel's theorem as:
\begin{equation}
  \label{eqn-l2-norm}
L_2(f) := \langle f, f \rangle^{1/2} = \left[\sum | A\mn |^2 \right]^{1/2},
\end{equation}
where $\langle a, b \rangle = \frac{1}{\pi r_0^2}\int_0^{r_0} a(r,\theta)
b^\star(r,\theta) r dr d\theta $ is the inner product for functions on the
disk.  The Sobolev $H_1(f)$ norm gives a measure of roughness,
or of how `wobbly' the function is across the disk:
\begin{align}
H_1(f) :=& \langle ( - \ell^2\nabla^2 + I) f, f \rangle ^{1/2}\notag\\
       =& \left[ \sum \Big(
         \frac{\ell^2\lambda\mn^2}{r_0^2} + 1\Big) |A\mn|^2 \right]^{1/2},
          \label{eqn-h1-norm}
\end{align}
where $\ell$ is a characteristic length scale introduced to maintain unit
consistency (we use $\ell = 1\fm$).  A variation on the Sobolev norm gives
the angular variation $M_1(f)$ which quantifies angular gradients,
\begin{equation}
  \label{eqn-m1-norm}
  M_1(f) :=  \langle \partial_{\theta}^2 f, f \rangle ^{1/2} 
         = \left[\sum m^2 |A\mn|^2\right]^{1/2}.
\end{equation}
We use these below to quantify roughness features of the events considered.
Note that although the individual coefficients $\{A\mn\}$ are not invariant
under translation or rotation of the coordinate system, the quantities
$L_2(f)$ and $H_1(f)$ {\bf are} invariant (and moreover scale simply with
changes in length scale $r_0$), while $M_1(f)$ is invariant to rotations.

To illustrate the usefulness of our proposed method, we apply it to three
example models: UrQMD, MC-Glauber and MC-KLN, which may be viewed as
representative of the models currently in use.

We consider a set of $100$ events generated by each code.  UrQMD includes
Boltzmann hadronic transport before the hydro begins, while the MC-Glauber
code includes simple streaming transport of the nucleons after interaction;
both of these will introduce added spatial fluctuation in the energy
density.  The Glauber code also includes KNO scaling of the multiplicity
fluctuations per binary collision.  For all models the initial condition
was computed in a $200$ point grid in the transverse plane at the center of
the collision along the beam axis.  To explore the centrality dependence of
the analysis we consider events at two impact parameters $b=2,7\fm$.  All
events are generated for Au+Au collisions at $\sqrt{s_{NN}} = 200\AGeV$.
To study the fluctuations generated by these events the ensemble averaged
event is computed for each model and subtracted from each event in the
ensemble before applying the decomposition.

We generated a series of ensembles of UrQMD events by averaging
successively larger samples of independent raw events together before
subtracting out the ensemble average.  See \cite{Petersen:2012qc} for more
details on this process and its influence on the ellipticity and
triangularity of the UrQMD initial state.

As the number $\Na$ of events over which we average increases, the scale of
the fluctuations will be diminished at approximate rate $1/\sqrt \Na$.
However this rescaling due to averaging events will not substantially change
the \emph {shape} of any particular fluctuation.  We have examined events
with $\Na=\{1, 2, 5, 10, 25\}$.  While only the $\Na=1$ case represents the
true UrQMD output the diminishing scale fluctuation in this sequence of
data sets will be used to illustrate different aspects of roughness in our
analysis.  Any statistic proposed as a measure of event fluctuations should
be invariant under this averaging procedure, which affects all frequencies
equally, and instead should be sensitive to smoothing procedures that
filter higher frequencies.

\begin{figure*}[htb]
  \centering
    \includegraphics[width=0.28\textwidth, clip, trim=0 0 0 5mm]{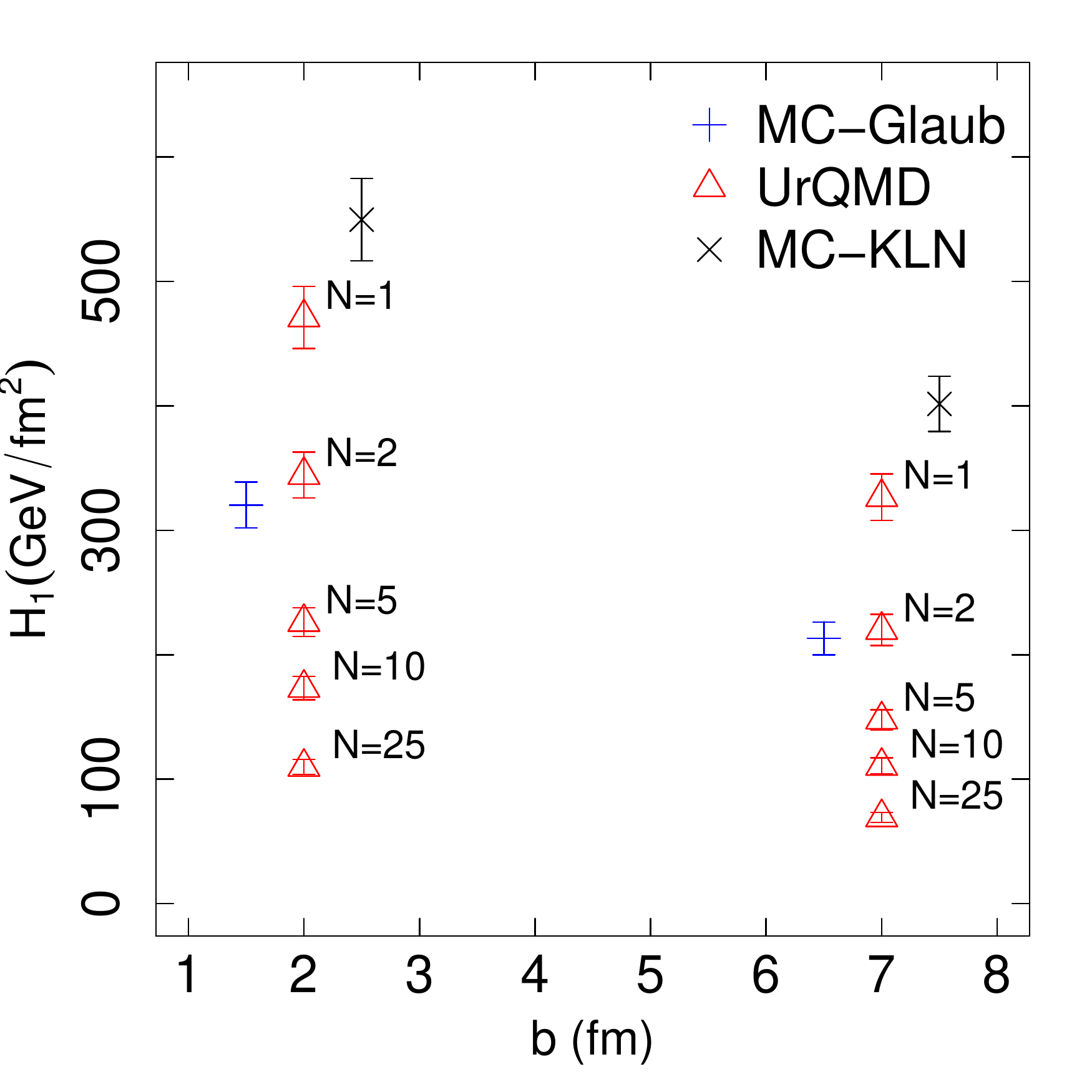}
    \includegraphics[width=0.28\textwidth, clip, trim=0 0 0 5mm]{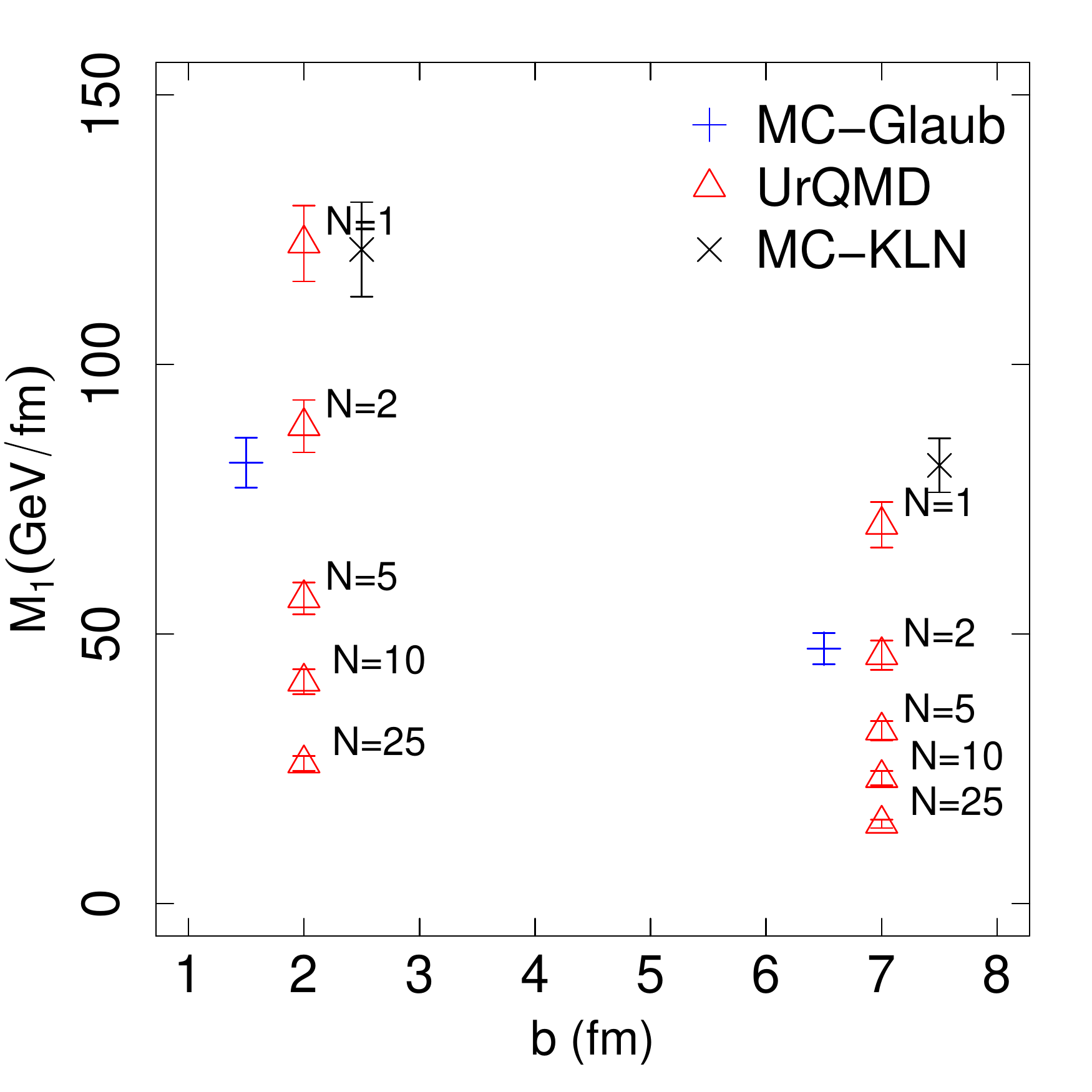}
    \includegraphics[width=0.28\textwidth, clip, trim=0 0 0 5mm]{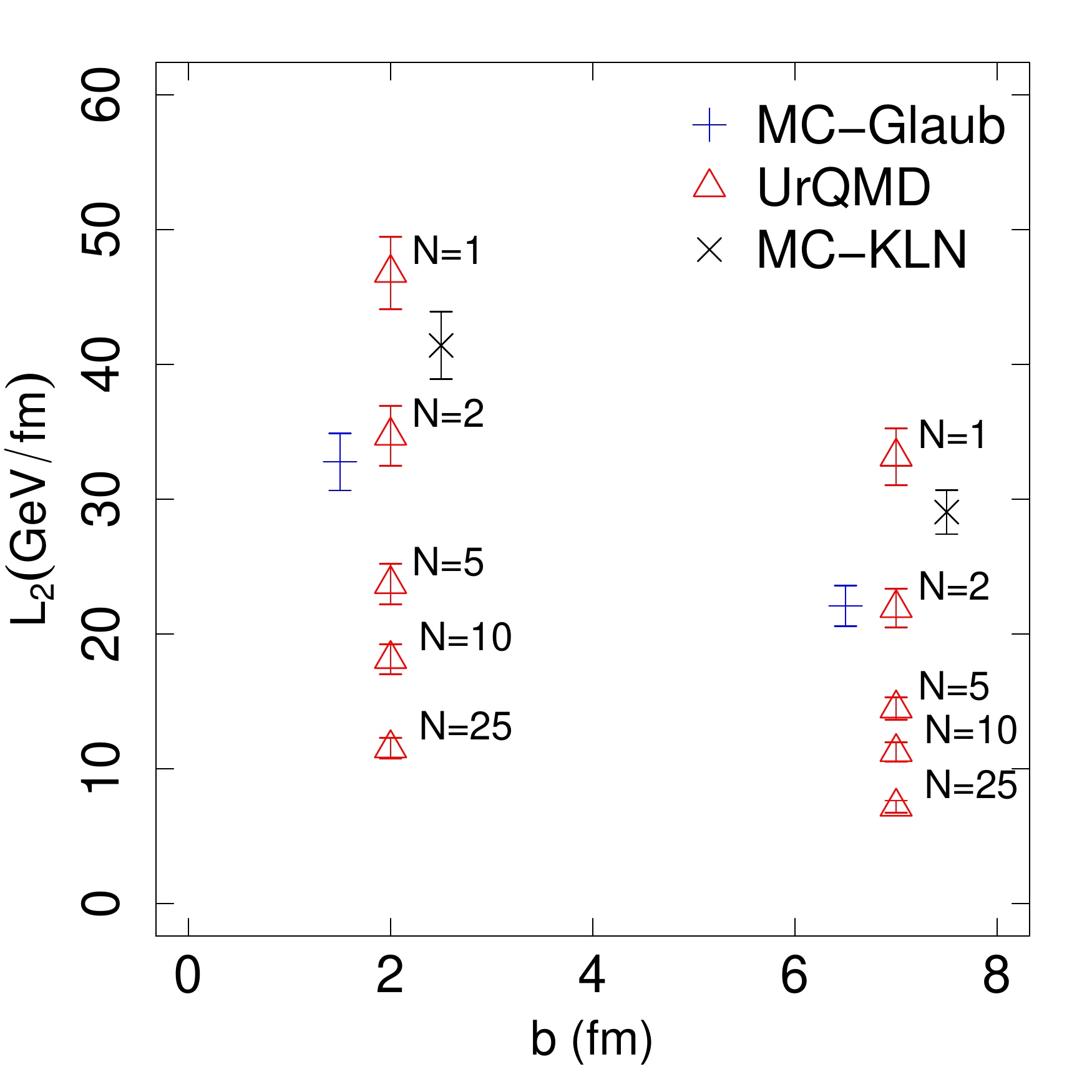}
    \caption{(color online) The ensemble averages of the Sobolev norm $H_1$ (left) the
      angular variation $M_1$ (center) and the $L_2$ norm (right).  The
      labels on the UrQMD glyphs give the number of averagings used to
      generate the ensemble members.  The MC-Glauber and MC-KLN results are
      plotted with an artificial offset in $b$ to permit easier
      comparison.}
  \label{fig-norm-compare}
\end{figure*}

In \figref{fig-norm-compare} the ensemble averages of each quantity ($H_1$,
$M_1$, $L_2$) are plotted for each set of events.  Considering first the
UrQMD results it is clear that each tends to zero as $\Na$ increases as
expected, confirming that each is quantifying fluctuations.  Fluctuations
are lower at the larger impact parameter $b$ but the relative ordering of
the models is preserved.  For all norms we see good agreement between the
$\Na=2$ UrQMD events and the MC-Glauber code.

The $L_2$ norm reflects the mean square fluctuation, the $H_1$ norm in
addition reflects the mean square spatial \emph{gradient} of the
fluctuations, and the $M_1$ semi-norm quantifies mean square angular
variation.  The MC-KLN results show the largest $H_1$ values while having
$M_1$ values comparable to UrQMD $\Na=1$, the MC-KLN $L_2$ is somewhat less
than UrQMD.  The relative ordering of the Glauber and UrQMD events is the
same across all norms.  The UrQMD and Glauber events are rather similar in
the scale and nature of their fluctuations, UrQMD produces slightly larger
fluctuations.  The MC-KLN model produces fluctuations on a scale comparable
to a hypothetical UrQMD $\Na=3/2$.  However the large $H_1$ and comparable
$M_1$ indicates that these events exhibit larger radial gradients than the
other models.  This may be attributed to the very rapid spatial falloff of
the gluon density near the edges of the nucleons.


The norms defined and plotted above, however informative, are
clearly quite sensitive to the overall scale of the fluctuations.  This is
shown most clearly in the separation over $\Na$ for the UrQMD results in
\figref{fig-norm-compare}.  We propose the following quantity (the
``roughness ratio'') as an overall scale invariant measure of the roughness
in an event:
\begin{align}
  \cR^2 &= \frac{H_1^2}{L_2^2} - 1
        = \frac{\langle -\ell^2\nabla^2 f,f\rangle}
               {\langle f,f\rangle}\label{eqn:norm-rough}
        &=\frac{\ell^2\sum \lambda\mn^2|A\mn|^2}{r_0^2\sum |A\mn|^2},
\end{align}
a weighted average of the scale-free squared characteristic inverse lengths
$(\lambda\mn\ell/r_0)^2$, weighted by the squared coefficients $|A\mn|^2$.
This is invariant to rescaling of $f$ or to Euclidean transformation, but
smoothing operations that preferentially reduce high-frequency components
of $f$ will reduce $\cR^2$ (but never below its minimum of
$(\lambda_{0,1}\ell/r_0)^2\approx 0.05783$). 

To illustrate that $\cR$ truly measures the degree of roughness in an event
we have have successively applied a finite difference smoothing operator (a
five point Laplacian stencil) to the event shown in \figref
{fig-decomp-example}, in effect melting the event.  We present the $H_1$,
$L_2$ and $\cR$ measures resulting from this process in \tabref
{tab-norm-compare}.  All of the norms decrease with the number of smoothing
iterations, however as we have shown above the $H_1$ and $L_2$ norms do not
exhibit an overall scale invariance which is desirable in an explicit
measure of roughness.  The $L_2$ norm gives the most natural measure of the
overall scale of fluctuations in an event.  The roughness ratio $\cR$
provides an explicitly scale invariant measure of the gradients, or
roughness, in an event.

\begin{table}[h]
  \centering
    \begin{tabular}{rrrr}
\multicolumn{1}{c}{$\Nf$}&
\multicolumn{1}{c}{$H_1^2$}&
\multicolumn{1}{c}{$L_{2}^2$}&
\multicolumn{1}{c}{$\cR^2$}\\
\hline
 0.00 & 3855.68 & 1754.63 & 1.20 \\ 
 32.00 & 3275.55 & 1547.44 & 1.12 \\ 
 128.00 & 2159.49 & 1108.36 & 0.95 \\ 
 1024.00 & 244.82 & 173.38 & 0.41 \\ 
 4096.00 & 27.87 & 24.80 & 0.12 \\ 
\end{tabular}
\caption{The $H_1^2$, $L_2^2$ and $\cR^2$ measures of a typical UrQMD event as a
  function of the number $\Nf$ of filtering passes, using a five-point
  finite difference smoothing filter.  The event we smoothed is shown in
  \figref{fig-decomp-example}.}
\label{tab-norm-compare}
\end{table}

\begin{figure*}[htb]
  \centering
    \includegraphics[width=0.3\textwidth]{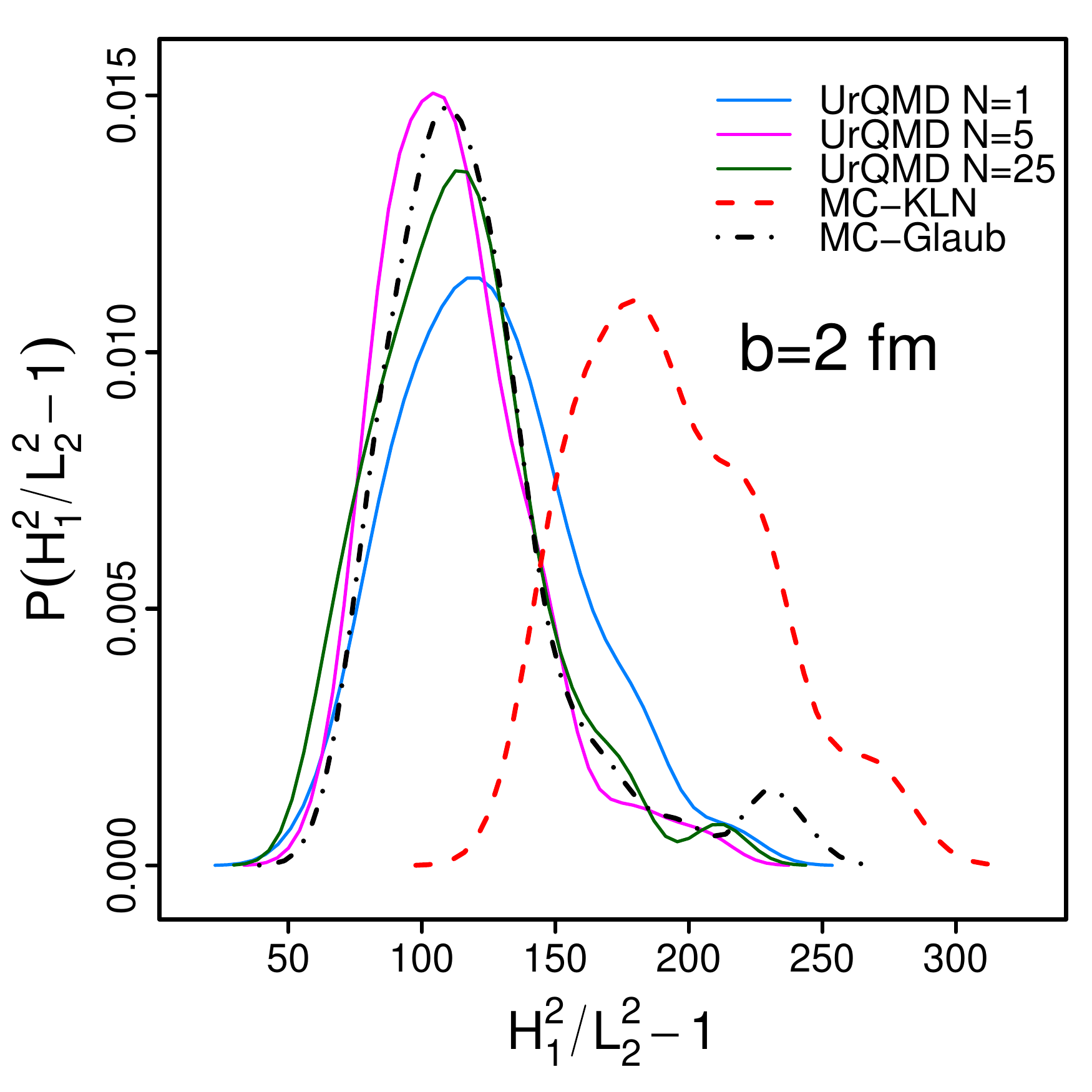}
    \includegraphics[width=0.3\textwidth]{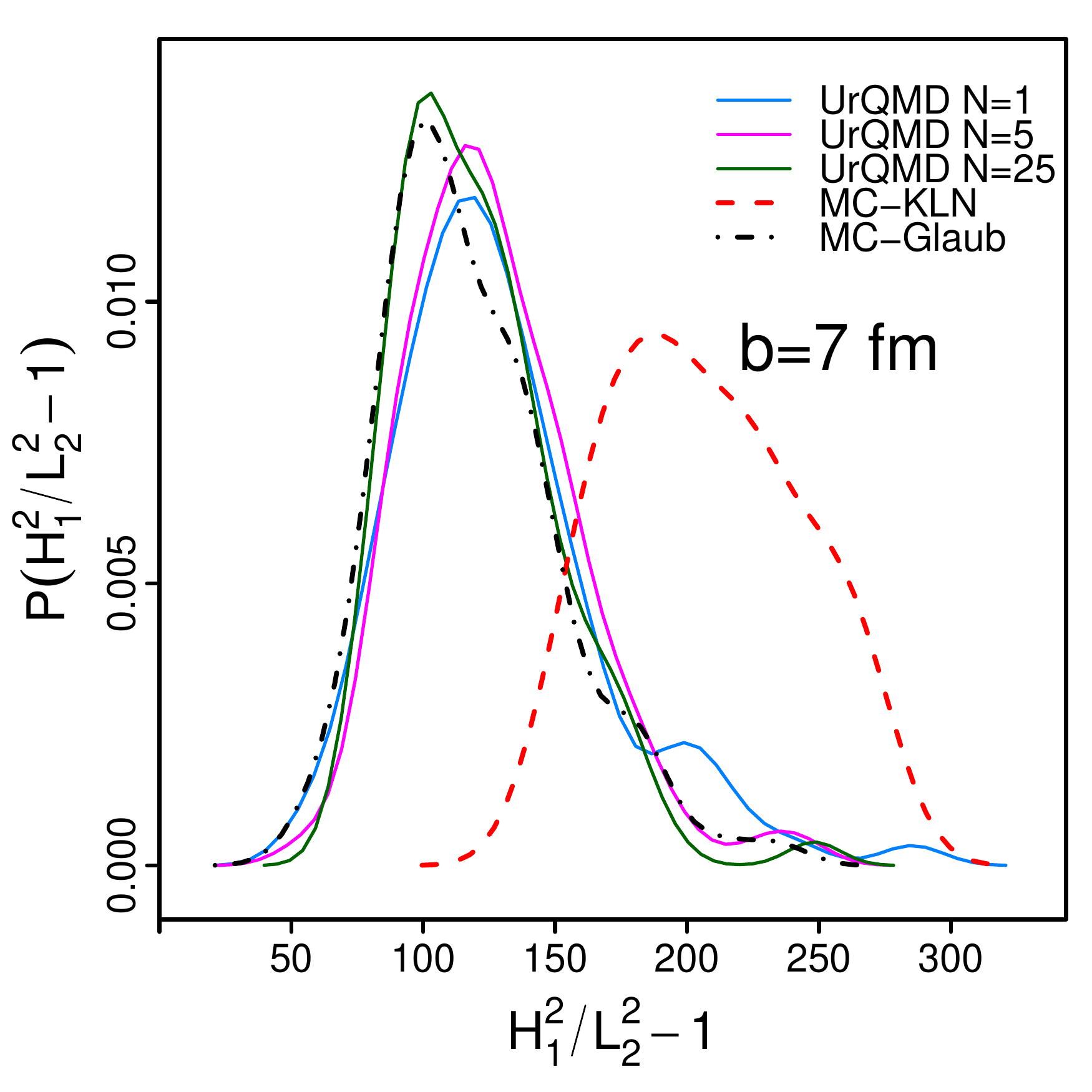}
    \caption{(color online) The distribution of $\cR^{2}$ (left, right) for UrQMD
      $\Na=\{1, 5, 25\}$, MC-KLN and MC-Glauber.  The left figure shows
      events at $b=2\fm$, the right figure shows $b=7\fm$.}
\label{fig-norm-rough-dist}
\end{figure*}
In \figref{fig-norm-rough-dist} we show the distribution of the roughness
ratio $\cR$ for UrQMD, MC-Glauber and KLN events at two centralities.  It
is invariant to the overall scale of the fluctuations in the event, the
curves for each of the $\Na$ UrQMD classes fall neatly on top of each
other.  Further we observe that the MC-Glauber and UrQMD events have very
similar distributions, this is reasonable given that the UrQMD events
originate from sampling Woods-Saxon profiles as well.  The distributions
for the KLN events clearly separate from the Glauber curves.  Given the
scale invariance of this quantity we conclude that the shape and extent of
the fluctuations in KLN events is fundamentally different from that in
equivalent Glauber events.


We have presented a new method for characterizing the fluctuations in the initial state of
heavy ion collisions.  The method is simple and general, it can be applied as easily to
theoretical models as to the output of event generators.  The $\cR$ measure we have
introduced is invariant under changes in coordinates, rotations and the overall scale of
the distributions.  We have shown the ability of this measure to quantify the broad
differences among the physical models we considered.  The radial information included
provides additional insights into the nature of fluctuations which are not readily
attainable by considering quantities derived from angular decompositions alone.


In future work we will examine how this measure passes through the hydrodynamical
evolution to the hadronic final-state of the collision.  Even if the observables developed
here cannot be measured in detectors they provide a useful basis for apples-to-apples
comparison of initial state models.

The authors are happy to provide sample analysis code to interested parties. 
  The authors would like to thank G.Y.  Qin for providing the MC-Glauber
  code.  This work was supported by U.S.\ Department of Energy grant
  DE-FG02-05ER41367 with computing resources from the OSG EngageVo funded
  by NSF award 075335, by NSF grants DMS--1228317 and PHY--0941373, and by
  NASA AISR grant NNX09AK60G.  Any opinions, findings, and conclusions or
  recommendations expressed in this material are those of the authors and
  do not necessarily reflect the views of the NSF or NASA.
  H.P. acknowledges support by the Helmholtz-Gemeinschaft for a HYIG grant
  VH-NG-822.

\appendix\section{Angular Decomposition}

To connect our analysis with commonly used quantities we quote in \tabref
{tab-epsilon-compare} the values of $\epsilon_2$ and $\epsilon_3$ computed
for events without ensemble average subtraction.

\begin{table}[h]
  \centering
    \begin{tabular}{l  c c c}
 & $b$ (fm) & $ E[\langle \epsilon_2 \rangle]$ & $ E[\langle \epsilon_3 \rangle]$ \\
\hline
UrQMD, $\Na=1$ & 2 & $0.096 \pm 0.005$ & $0.079 \pm 0.004$ \\
MC-KLN       & 2 & $0.084 \pm 0.005$ & $0.046 \pm 0.003$\\
MC-GLAUB     & 2 & $0.089 \pm 0.005$ & $0.070 \pm 0.004$\\
UrQMD, $\Na=1$ & 7 & $0.271 \pm 0.010$ & $0.117 \pm 0.006$ \\
MC-KLN       & 7 & $0.343 \pm 0.010$ & $0.111 \pm 0.006$\\
MC-GLAUB     & 7 & $0.231 \pm 0.009$ & $0.099 \pm 0.006$\\
\end{tabular}
\caption{The ensemble average values of $\langle \epsilon_2 \rangle$ and $
  \langle \epsilon_3 \rangle$ for each of the models at impact parameters
  $b=2,7\fm$.}
\label{tab-epsilon-compare}
\end{table}

\section*{References}
\bibliographystyle{hunsrt}
\bibliography{./fourier-cites-noarxiv}{}
\end{document}